\documentclass[12pt,a4paper]{article}
\usepackage{amsmath,amssymb}
\usepackage{graphicx}
\usepackage{float}
\usepackage{cite}
\usepackage{url}
\usepackage{geometry}
\title{Response of Elliptical Scatterer Due to Perfect Magnetic Material}
\author{
    Waqas Ahmed\textsuperscript{1}, 
    Ahsan Illahi\textsuperscript{2},
    Asma\textsuperscript{3}, 
    }
\date{}

\begin{document}

\maketitle

\begin{center}
    \textsuperscript{1}Center for Fundamental Physics, School of Artificial Intelligence, \\Hubei Polytechnic University, Huangshi 435003, China    \\

    \textsuperscript{2,3}Research in Modeling \& Simulation (RIMS) Group, Department of \\
    Physics, COMSATS University Islamabad, Islamabad Pakistan.
\end{center}

\begin{center}
\end{center}


\begin{abstract}
The effects on the bistatic echo width of an elliptical cylinder due to a perfect magnetic material are reported in this article. The configuration is analyzed using the separation of variables method and Mathieu functions. In this approach, the structural geometry is illuminated by an electromagnetic field. Radial and angular Mathieu functions have been used in the formulation. Notably, the maxima of the scattered elliptic transfer electric mode ($\theta = 180^{\circ}$) are much higher than those of the scattered transfer magnetic mode, comparable to terms $\theta = 120^{\circ}$ and $\theta = 240^{\circ}$, respectively. It can be observed that an increase in the in-plane radial component leads to the linearity principle for the transfer electric mode, while non-linear behavior is investigated for the elliptic transfer magnetic mode. Therefore, the unidirectional bistatic echo width is subject to non-directional behavior. These analogous results may have applications in the fields of optics, meteorology, acoustics, radio astronomy, collision physics, and other disciplines where wave scattering phenomena play a crucial role. Furthermore, the findings of this study contribute to the fundamental understanding of electromagnetic interactions with complex geometries and materials.
\end{abstract}

\section{Introduction}
Scattering problems involving homogeneous and inhomogeneous elliptical cylinders hold significant importance due to their wide applications in electromagnetics [1-25], including radar sensing, computer-generated imagery, medical ultrasound, meteorology, acoustics, and radio astronomy, among others. The utilization of Mathieu functions aids in deriving scattered field expressions as functions of elliptical cylinder coordinates through expansion coefficients of closed-form expressions [1]. Scattering occurs when radiation is deviated from its original path due to non-uniformities in the propagation medium. The field pattern from a dielectric-coated elliptic cylinder has been studied [6] and a spatially uniform coated impedance elliptical cylinder scattering is investigated [7]. Exact solutions for multilayered dielectric elliptical cylinders are found using recursive relation formulations [8], and scattering theory regarding exact solutions for planar waves on iso-refractive cylinders is explored [10].

An intriguing observation is that when dielectric permittivity and magnetic permeability values of an elliptical cylinder are mismatched, the incident electromagnetic mode is represented by a single mode of Mathieu functions whereas the scattered field is represented by modes of different orders  Mathieu functions [11, 12]. This effect is demonstrated when an electric current source representing the single-mode Mathieu function is excited by an electromagnetic field [1-14]. This is crucial fpr understanding the behaviour of the system when plane wave excitation is commonly used. Modes in plane waves are infinite in circular and elliptical coordinates, making calculations for elliptically shaped objects complex due to coupled elliptical modes for scattered fields.

There's growing interest in electromagnetic wave scattering by realistic geometries such as spheres, circular cylinders, and elliptical cylinders, leading to considerable attention in the literature [5]. Elliptic cylinders offer generalized geometry, capable of producing various cross-sectional cylindrical objects by altering the axial ratio. The wave equation is separable in an elliptical coordinate system, allowing closed-form solutions for problems involving ellipses.

Recent years have seen significant research on scattering by perfect magnetic objects, due to applications in engineering and theoretical studies. Perfect Magnetic Conductors (PMC) attract researchers in electromagnetic tunneling, absorbers, and antennas. Unlike Perfect Electric Conductors (PEC), PMC doesn't alter the electric field's phase upon reflection but reverses the magnetic field's phase [1].

This paper is invaluable for scholars exploring scattering from small objects. Adjusting the elliptical geometry's eccentricity enables constructing ellipses and objects with thin or flattened scattering profiles, such as radars, planks of wood, absorbing-coated aircraft wings, fingers [17], or wrists. The analysis here focuses on the scattering problem of plane waves involving an infinitely long perfectly magnetic elliptic cylinders, considering both TE and TM polarizations and arbitrary angles of incidence and axial ratios. The boundary condition $\widehat{\mathbf{n}}\times \mathbf{H} = 0$, on the PMC surface is met at low frequencies using high-$\mu$ materials. For magnetostatics, at the interface of air and a high-permeability material, the magnetic flux is normal to the surface as required for PMC material. At high frequencies, finding materials with high permeability is challenging, making it feasible to mimic the PMC condition on a surface where no surface current can flow.

\section{Analytical Model}
A Perfect Magnetic Conductor (PMC) is an idealized medium with infinite magnetic conductivity. While perfect magnetic conductors do not occur naturally, the concept is useful when magnetic resistances are negligible compared to other effects. At the surface of a PMC object, the required boundary conditions are as follows:

\begin{equation}
\widehat{\mathbf{n}}\times \mathbf{H} = 0,\quad\quad \widehat{\mathbf{n}}\cdot \mathbf{B} = 0
\end{equation}

Here, $\mathbf{H}$, $\mathbf{B}$ and $\mathbf{n}$ represent the magnetic fields, magnetic flux densities, and the unit normal to the surface boundary, respectively.

Consider a uniform plane electromagnetic (EM) wave incident on an infinitely long PMC elliptic cylinder with semi-major axis '$a$' and semi-minor axis '$b$', where the $z$-axis is the cylinder axis. Illustrated in Figure 1, the plane wave approaches at an angle $\phi_{i}$ relative to the positive $x$-axis of the Cartesian coordinate system. For analysis, it's advantageous to present the problem in elliptic coordinates $u$ and $v$, defined as $x = F\cosh u\cos v$, $y = F\sinh u\sin v$, with the semi-focal length of the ellipse as $F$. Throughout this study, the time-dependent factor $\exp(j\omega t)$ is considered, where $\omega$ is the angular frequency.

\subsection{TM Polarization}
\begin{figure}[H]
\centering
\includegraphics[width=0.5\textwidth]{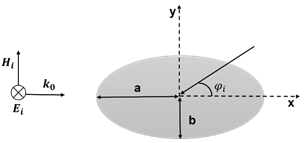}
\caption{Configuration for TM polarization}
\end{figure}




The longitudinal component of the incident TM-polarized field may be represented as
\[
E_{z}^{(i)} = \exp\!\left[-jk\big(\cos\varphi_{i} + \sin\varphi_{i}\big)\right],
\]
where $k$ denotes the wavenumber corresponding to the wavelength $\lambda$, and $\varphi_{i}$ specifies the angle at which the plane wave arrives.  

For an elliptic cylindrical coordinate system, the associated vector wave functions are built from the scalar Mathieu functions.  
The electric-type function $\vec{N}$, and the magnetic-type function $\vec{M}=k^{-1}\nabla\times\vec{N}$, can be written in the form

\begin{equation}
\vec{N}_{qm}^{(i)}(c,\xi,\eta)
 = R_{qm}^{(i)}(c,\xi)\, S_{qm}(c,\eta)\, \hat{\mathbf{z}},
\end{equation}

\begin{equation}
kh\,\vec{M}_{qm}^{(i)}
 = R_{qm}^{(i)}(c,\xi)\, S_{qm}'(c,\eta)\,\hat{\mathbf{u}}
 - R_{qm}^{(i)\,'}(c,\xi)\, S_{qm}(c,\eta)\,\hat{\mathbf{v}} ,
\end{equation}

where $R_{qm}^{(i)}$ and $S_{qm}$ denote the radial and angular Mathieu functions, respectively, primes indicate differentiation with respect to the argument, and $h$ is the scale factor in elliptic coordinates.


\begin{equation}
E^{i} = \sum_{m = 0}^{\infty}{A_{em}N_{em}^{(1)} + \sum_{m = 1}^{\infty}{A_{om}N_{om}^{(1)}}}
\end{equation}

$A_{em}$ and $A_{om}$ are the coefficient of expansion [9]. The scattered field can be assumed of the form as follows.

\begin{equation}
E^{s} = \sum_{m = 0}^{\infty}{\left\{ B_{em}N_{em}^{(4)} \right\} + \sum_{m = 1}^{\infty}{\{ B_{om}N_{om}^{(4)}\}}}
\end{equation}

In eq 5 $B_{em},B_{om}$ are unknown coefficients of expansion. Similarly, the incident and scattered magnetic field can be evaluated using eq (4), eq (5), and the Maxwell Eq as:
\begin{equation}
H^{i} = \frac{j}{Z} \left[ \sum_{m = 0}^{\infty} A_{em} M_{em}^{(1)} + \sum_{m = 1}^{\infty} A_{om} M_{om}^{(1)} \right]
\end{equation}

\begin{equation}
H^{s} = \frac{j}{Z}\left[ \sum_{m = 0}^{\infty}{\{ C_{em}}N_{em}^{(4)}\} + \sum_{m = 1}^{\infty}{\{ C_{om}N_{om}^{(4)}\}} \right]
\end{equation}

In eq 7 $C_{em}, C_{om}$ are the unknown coefficients of expansion and $Z$ is the impedance of wave in free space. By using Tangential condition at the interface of cylinder to obtain unknown coefficients of expansion as:

\begin{equation}
\left[H_{z}^{i} + H_{z}^{s}\right] = 0
\end{equation}

\begin{equation}
\left[H_{v}^{i} + H_{v}^{s}\right] = 0
\end{equation}

Applying the orthogonal property of Mathieu functions and substituting value of electric and magnetic field in above equations yield:

\begin{equation}
B_{qm} = - \frac{R_{qm}^{(1)'}(c, \xi_{s})R_{qm}^{(4)}(c, \xi_{s})}{R_{qm}^{(4)}\left( c, \xi_{s} \right)R_{qm}^{(4)'}(c, \xi_{s})}A_{qm}
\end{equation}

The normal boundary condition (1) is satisfied by the incident fields and scattered fields obtained using (10).

\subsection{TE Polarization}
\begin{figure}[H]
\centering
\includegraphics[width=0.4\textwidth]{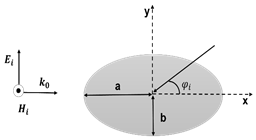}
\caption{Configuration for TE polarization}
\end{figure}

Using duality, eq. 4 to eq (7) are used to obtain the expressions for the incident and scattered EM fields for TE case.  The coefficient of expansion in eq (10) is obtained by satisfying boundary conditions on the interface of cylinder.

\subsection{Far Field}
Echo width is one of the most important parameters in scattering, it is obtained by calculating scattered field in far zone. For incident transverse magnetic incident wave, the scattered components in far zone ($\xi \rightarrow \infty$) are obtained using asymptotic expressions of $R_{qm}^{(4)}(c,\xi)$ and $R_{qm}^{(4)'}(c,\xi)$. Considering $\tau = \cos\phi$, the electric and magnetic field components are,

\begin{equation}
E_{z}^{s} = \sqrt{\frac{j}{k\rho}} \, e^{-jk\rho} \left[ 
\sum_{m = 0}^{\infty} j^{m} B_{em} S_{em}(c,\tau) + 
\sum_{m = 1}^{\infty} j^{m} B_{om} S_{om}(c,\tau) 
\right]
\end{equation}

\begin{equation}
E_{\phi}^{s} = \sqrt{\frac{j}{k\rho}} \, e^{-jk\rho} \left[ 
\sum_{m = 0}^{\infty} j^{m} C_{em} S_{em}(c,\tau) + 
\sum_{m = 1}^{\infty} j^{m} C_{om} S_{om}(c,\tau) 
\right]
\end{equation}

\begin{equation}
H_{\phi}^{s} = - \frac{E_{z}^{s}}{Z}
\end{equation}

\begin{equation}
H_{z}^{s} = \frac{E_{\phi}^{s}}{Z}
\end{equation}

The bistatic echowith can be defined as follow:

\begin{equation}
\sigma = \lim_{\rho \rightarrow \infty}{2\rho\pi\frac{\Re\left[\left( E^{s} \times H^{s*} \right)\cdot\widehat{\rho}\right]}{\Re\left[\left( E^{i} \times H^{i*} \right)\cdot\widehat{\rho}\right]}}
\end{equation}

where $\widehat{\rho}$ denotes the unit vector in the increasing direction of $\rho$ and asterisk represents the complex conjugate. After substituting scattered and incident fields in (15), the (normalized) bistatic echo width is obtained:

\begin{equation}
\begin{aligned}
\frac{\sigma}{\lambda}
&=
\sum_{m=0}^{\infty}\sum_{n=0}^{\infty}
j^{m}(-j)^{n}\,
\bigl(B_{em}S_{em}(c,\tau)+B_{om}S_{om}(c,\tau)\bigr)
\bigl(B_{en}S_{en}(c,\tau)+B_{on}S_{on}(c,\tau)\bigr)^{*}
\\[6pt]
&\quad+
\sum_{m=0}^{\infty}\sum_{n=0}^{\infty}
j^{m}(-j)^{n}\,
\bigl(C_{em}S_{em}(c,\tau)+C_{om}S_{om}(c,\tau)\bigr)
\bigl(C_{en}S_{en}(c,\tau)+C_{on}S_{on}(c,\tau)\bigr)^{*}.
\end{aligned}
\end{equation}

By substituting $\phi = \phi_{i}$ in (16) the normalized backscattering width can be obtained.

\section{Numerical Results and Discussion}
The approach described above was used to investigate the normalized scattering cross-section of elliptic PMC cylinder for both TE and TM polarized incident waves. Results are presented for different axial ratio and by varying the angle of incidence. To validate our formulation for simulation, the normalized backscattering widths have been calculated first for PMC elliptic cylinders excited by normally incident plane wave when axial ratio is 2 and length of semi-major axis is $\frac{\lambda}{2}$. Normalized scattering width $\frac{\sigma}{\lambda}$ in Fig. 3 compared with [2] verifies the accuracy of analysis.

\begin{figure}[H]
\centering
\includegraphics[width=0.45\textwidth]{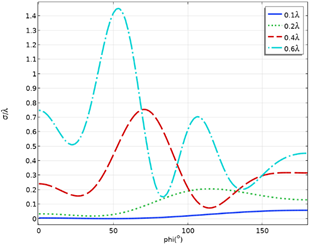}
\caption{Validation and comparison results}
\end{figure}

The magnitude of scattering of TE wave is large as compared to TM wave and maximum scattering occurs at $180^o$ whereas, for TM maximum scattering is observed at $120^o$ and $240^o$. The normalized backscattering width ($\sigma/\lambda$) for an elliptic PMC cylinder of axial ratio 2 is plotted against $ka$ in Fig. 4. 

\begin{figure}[H]
\centering
\includegraphics[width=0.45\textwidth]{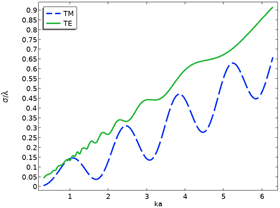}
\caption{Validation and comparison results}
\end{figure} 

Here we have observed an increase in the scattering width with increasing length of semi major axis. For TE incident wave the relation become linear at large $ka$ values whereas for TM the variation is nonlinear.
\begin{figure}[H]
\centering
\includegraphics[width=0.45\textwidth]{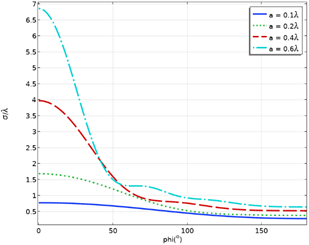}
\caption{Backscattering width variation}
\end{figure}
Plots of the bistatic $\frac{\sigma_{2-D}}{\lambda}$ are computed in Fig. 5 for PMC elliptic cylinder with semi-major radii of $a = 0.1\lambda, 0.2\lambda, 0.4\lambda,$ and $0.6\lambda$ and axial ratio 2.
Plots of the bistatic $\frac{\sigma_{2-D}}{\lambda}$ are computed in Fig. 5 for PMC elliptic cylinder with semi-major radii of $a = 0.1\lambda, 0.2\lambda, 0.4\lambda,$ and $0.6\lambda$ and axial ratio 2.
\begin{figure}[H]
\centering
\includegraphics[width=0.45\textwidth]{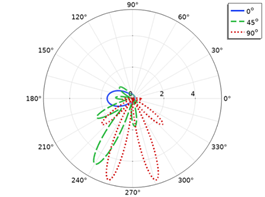}
\includegraphics[width=0.45\textwidth]{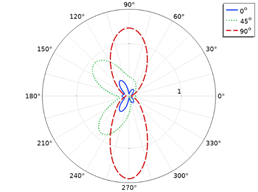}
\caption{Backscattering width variation}
\end{figure}


 For an axially incident TE wave as predicted there is an increase in scattering width with the size of cylinder and echo width is maximum for all cylinders at $\phi = 0^o$.

Also, for TM wave scattering increases with the size of cylinder but it is not unidirectional and maximum scattering moves toward lower angles at large sizes.

2d RCS is plotted for the PMC elliptic cylinder considered in Fig. 3, when plane wave is incident at $0^o$, $45^o$ and $90^o$ shown in Fig. 6. Fig. 6(a) is the variation for TE wave, when $\phi_{i} = 0^o$ maximum scattering occurs at $180^o$, for $\phi_{i} = 45^o$ along with several side lobes main lobe is present at $240^o$ whereas, maximum scattering width is observed for $\phi_{i} = 90^o$, in this case two main lobes are present at $255^o$ and $285^o$. Similarly, for TM wave maximum NBSW with broader lobes are observed at $\phi_{i} = 90$.
\begin{figure}[H]
\centering
\includegraphics[width=0.45\textwidth]{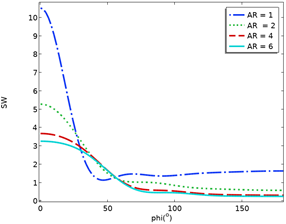}
\caption{Axial ratio variations}
\end{figure}
 In Fig. 7 and 8, the variation of the NBSCW is shown with $\phi_{i} = 0^o$ for PMC elliptic cylinders of semi-major axis length $\lambda/2$, and four different axial ratios, are shown . Due to symmetry the axis have been cut to $180^o$. For TE case, maximum scattering occurs for lower axial ratios (AR), peak width of all patterns is at $\phi = 0^o$, and they become broader with AR. When incident wave is TM, the scattering become more directional with axial ratio, maximum scattering is observed at $60^o$ when AR increase from 2 to 6 and peak reaches $0^o$ for AR = 1.

\begin{figure}[H]
\centering
\includegraphics[width=0.45\textwidth]{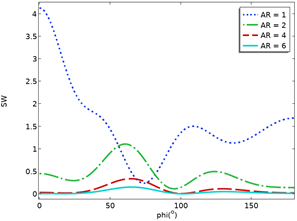}
\caption{Axial ratio variations}
\end{figure}

\section{Conclusions}
This paper gives an insight into the scattering behavior of PMC elliptic cylinder when illuminated with transverse electric and transverse magnetic plane waves. The formulation is purely analytic and, in this way, unique. In earlier work published by different scientists, the problem is usually handled by the duality principle. The increasing trend in the synthesis of magnetic materials requires this type of sole formulation. Bistatic radar cross section is analyzed by varying incident angle, size and axial ratio of cylinder. These results are important for validation purpose as most of the real scatterers can be presented by elliptic cylinders. Furthermore, PMC boundaries and their interaction with EM field are very important in commercial software and EM solvers for symmetry purpose. The results presented in the current paper are therefore more important for comparison purposes by the end users of the available commercial EM solvers.

\end{document}